\documentclass[fleqn,usenatbib]{mnras}

\usepackage{newtxtext,newtxmath}

\usepackage[T1]{fontenc}

\DeclareRobustCommand{\VAN}[3]{#2}
\let\VANthebibliography\thebibliography
\def\thebibliography{\DeclareRobustCommand{\VAN}[3]{##3}\VANthebibliography}

\usepackage{graphicx}	
\usepackage{amsmath}	
\usepackage{graphicx, times}             
\usepackage{natbib}
\usepackage{soul}
\soulregister\citealt7
\soulregister\citet7
\soulregister\ref7
\usepackage{verbatim}
\usepackage{subfig}
\usepackage{stfloats}
\everymath{\displaystyle}

\title[Systems Generating `Oumuamua-Like Asteroids]{Configuration of Single Giant Planet Systems Generating `Oumuamua-Like Interstellar Asteroids}

\author[Xi-Ling Zheng \& Ji-Lin Zhou]{
Xi-Ling Zheng,$^{1,2}$\thanks{E-mail: zhengxl@smail.nju.edu.cn}
Ji-Lin Zhou,$^{1,2}$\thanks{E-mail: zhoujl@nju.edu.cn}
\\
$^{1}$School of Astronomy and Space Science, Nanjing University, Nanjing 210023, China\\
$^{2}$Key Laboratory of Modern Astronomy and Astrophysics, Ministry of Education, Nanjing 210023, China
}

\date{Accepted 2025 February 4. Received 2025 January 30; in original form 2024 December 10}

\pubyear{2025}

\begin{document}
\label{firstpage}
\pagerange{\pageref{firstpage}--\pageref{lastpage}}
\maketitle

\begin{abstract}
The first discovered interstellar small object, `Oumuamua (1I/2017 U1), presents unique physical properties of extremely elongated geometric shape and dual characteristics of an asteroid and a comet. These properties suggest a possible origin through tidal fragmentation, which posits that `Oumuamua was produced through intensive tidal fragmentation during a close encounter with a star or a white dwarf, resulting in its shape and ejection from its natal system. According to this mechanism, a high initial orbit eccentricity and a small pericentre of the parent body are necessary to produce `Oumuamua-like objects. To verify whether this mechanism can occur in single giant planet systems, we conduct long-term numerical simulations of systems with a low-mass ($0.5M_\odot$) host star and a giant planet in this study. We determine that an eccentric orbit ($e_\text{p}\sim0.2$) and a Jupiter-mass ($M_\text{p}\sim M_\text{J}$) of the planet appears to be optimal to generate sufficient perturbations for the production of ‘Oumuamua-like objects. When the planetary semi-major axis $a_\text{p}$ increases, the proportion of planetesimals ejected beyond the system $P(\text{ej})$ increases accordingly, while the possibilities of ejected planetesimals undergoing stellar tidal fragmentation $P(\text{tidal}|\text{ej})$ remains relatively constant at $\sim0.6\%$. Focusing on stellar tidal fragmentation alone, the ratio of extremely elongated interstellar objects to all interstellar objects is $P_\text{e}\sim3\%$. 

\end{abstract}




\section{Introduction}

\indent\indent With a velocity of $25\,\mathrm{km\,s^{-1}}$ at infinity and an orbital eccentricity reached $1.1956\pm0.0006$, `Oumuamua (1I/2017 U1) was confirmed as the first interstellar small object discovered within the Solar System (\citealt{Jewitt+2017}). `Oumuamua demonstrates a substantial brightness variation exceeding $2.5$ magnitudes over a brief interval of approximately $4\,\text{h}$ (\citealt{Meech+2017}; \citealt{Bannister+2017}; \citealt{Knight+2017}; \citealt{Bolin+2018}; `Oumuamua ISSI Team \citeyear{Team+2019}). The ratio of its long to short axes is inferred from the brightness variation up to $\ge 6:1$ (\citealt{McNeill+2018}; \citealt{Drahus+2018}), and later inferred by calculations $\sim 6:6:1$ (\citealt{Mashchenko+2019}). Few known objects within the solar system exhibit comparable luminosity variations and extreme axis ratio (\citealt{Warner+2009}; \citealt{Durech+2012}; \citealt{Meech+2017}). Additionally, `Oumuamua lacks observed cometary characteristics, its dry and rocky surface suggests an asteroidal nature, however, its non-gravitational acceleration implies comet-like properties (\citealt{Meech+2017}; \citealt{Bannister+2017}; \citealt{Micheli+2018}; \citealt{Fitzsimmons+2018}).

Since `Oumuamua was discovered, its perplexing properties have presented a challenge to researchers. A theory of tidal fragmentation, proposed by \citet{Zhang+2020} (Hereafter ZL20), provides a plausible explanation and effectively accounts for the existence of `Oumuamua-like objects. Utilizing high-precision numerical simulations, ZL20 examined the dynamics and thermodynamic evolution of the parent bodies (kilometre-sized planetesimals/comets or planet-sized bodies) during their close encounters with a star. Their findings indicate that the tidal forces from the star can shatter the parent bodies into elongated fragments, leading to their extreme axis ratios. Heat from the star and subsequent re-condensation render the fragments' structure highly cohesive, thereby preserving the stability of the elongated shapes. Concurrently, the star's thermal radiation heats the fragments' interiors, prompting the evaporation of volatile gases, which imparts spectral characteristics typically associated with asteroids, devoid of distinct cometary activity.

Previous work has also calculated the probability of planetesimals being tidally fragmented but mainly focused on planetary tidal fragmentation. \citet{Raymond+2020} is an example of the evolution in the Solar System, which specifically examined the different fates of ejected versus surviving planetesimals and found that ejected planetesimals experienced tidal fragmentation at more than twice the rate of surviving planetesimals (3.1\% versus 1.4\%). \citet{Raymond+2018} explored the possibility of `Oumuamua as a fragment of an ejected cometary planetesimal, indicating that for a solar-mass star with three giant planets on near-circular orbits and an outer disc of planetesimals, its dynamical evolution over $100$ to $200$ Myrs could produce an object similar to `Oumuamua through planetary tidal fragmentation. In contrast to previous studies, the focus of this work will be on planetesimals' tidal fragmentation by a star to follow the simulation results of ZL20.

\begin{figure}
    \centering
    \includegraphics[width=0.45\textwidth]{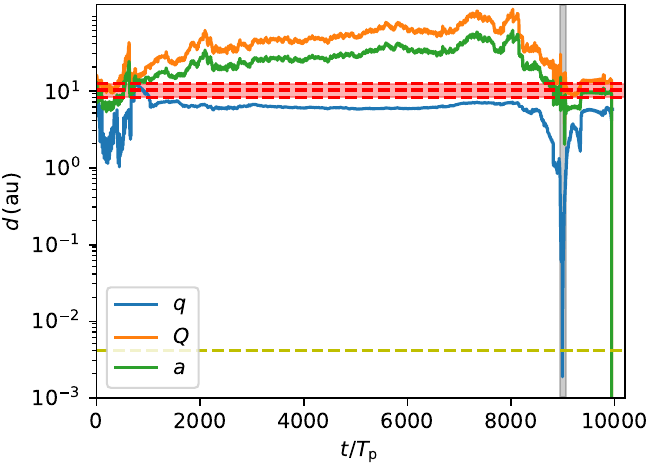}
    \caption{Evolution of an example planetesimal in a low-mass star ($M_\text{s}=0.5M_\odot$, $R_\text{s}=0.5R_\odot$) and cold Jupiter system ($a_\text{p}=10.4\,\text{au}$, $e_\text{p}=0.2$) with eventual $e>1$ and the minimum distance between the star and the planetesimal is less than the tidal fragmentation limit distance ($d_
    {\min}<d_{\text{str}}$). The blue, orange, and green solid lines represent the osculating perihelion, aphelion, and semi-major axis of the planetesimal during its evolution, respectively. The negative semi-major axis $a<0$ at $t\sim10^4T_\text{p}$ corresponds to hyperbolic orbit. The red dashed lines represent the perihelion, semi-major axis, and aphelion of the cold Jupiter's orbit. The gray area illustrates the periastron of the planetesimals gradually approaching the star, and its minimum distance from the star entering the tidal fragmentation limit of the star, then it is deduced to be tidally fragmented.
    }
    \label{example}
\end{figure}

While the tidal fragmentation theory of ZL20 offers a plausible explanation for `Oumuamua's formation, it imposes strict requirements on the initial eccentricity and pericentre of the parent body. Simulations suggest that near a host star with a mass of $0.5M_\odot$, parent bodies with sizes of $R=100\,\text{m}$ require an initial eccentricity of $1-e_0\approx10^{-6}$ to generate `Oumuamua-like fragments. This implies that if the parent body is a planetesimal, it has likely been subjected to secular perturbations by planets or other celestial bodies either before or after tidal fragmentation, which can excite the object's eccentricity to a significant value (\citealt{Zhou+2007}). To verify whether the probability of this mechanism occurring in single giant planet systems can justify the reasonableness of ‘Oumuamua’s detection, we conduct numerical simulations to investigate whether the perturbation by a giant planet can 1) excite the eccentricity of a planetesimal to a value $\gtrsim1$ and 2) provide the planetesimal with a small enough distance from the star to satisfy the requirement for tidal fragmentation. An example is depicted in Fig. \ref{example}.

In Section \ref{sect:2} of this study, we present our simulations of planetesimal evolution within the Sun-Jupiter system, including the evolution patterns of orbital elements and the distribution of minimum distances from the planetesimals to the host star during the evolution, and then use the results to estimate the likelihood of tidal fragmentation events. Section \ref{sect:3} expands our analysis to various exoplanet systems, applying the same methodology. The discussion of the simulation results is presented in Section \ref{sect:4}. Last, Section \ref{sect:5} offers a summary of our findings along with their broader implications.

\section{Evolution of Planetesimals in Sun-Jupiter System}
\label{sect:2}

\indent\indent We examine the influence of diffusion during the dynamical evolution of the Sun-Jupiter system over a time-scale of $10^6$ times of Jupiter's orbital period $T_\text{J}$. The simulations are conducted within the framework of the Sun-Jupiter-planetesimal restricted three-body system (under which the planetesimals are massless). Jupiter's orbit is assumed to be elliptical, with an eccentricity of $e_\text{J}=0.048$. Initial eccentricities and inclinations of the planetesimals are set to $e = 0.02$ and $i = 0.01$, while other orbital elements—mean anomalies, longitudes of perihelion, and arguments of perihelion—are randomized.

Planetesimals located within the region known as the chaotic zone face a substantial likelihood of experiencing close encounters with Jupiter. The inner and outer boundary ($a_\text{in}$, $a_\text{out}$) of the chaotic zone are given by \citet{Petrovich+2015}:
\begin{align}
\left.
\begin{aligned}
\frac{a_{\text{p}}\left(1-e_{\text{p}}\right)}{a_{\text {in }}}&=2.4\left(\frac{M_{\text{p}}}{M_{\text{s}}+M_{\text{p}}}\right)^{\frac{1}{3}}\left(\frac{a_{\text{p}}}{a_{\text {in }}}\right)^{\frac{1}{2}}+1.15 
\\
\frac{a_{\text {out }}}{a_{\text{p}}\left(1+e_{\text{p}}\right)}&=2.4\left(\frac{M_{\text{p}}}{M_{\text{s}}+M_{\text{p}}}\right)^{\frac{1}{3}}\left(\frac{a_{\text{out}}}{a_{\text{p}}}\right)^{\frac{1}{2}}+1.15
\end{aligned}
\right.
\label{chaotic}
\end{align}

where $a_\text{p}$, $e_\text{p}$ and $M_\text{p}$ represent the semi-major axis, eccentricity and mass of the planet, respectively. $M_\text{s}$ denotes the mass of the star. The boundaries of the chaotic zone in the Sun-Jupiter system are calculated to be $a_\text{in}=0.661a_\text{J}=3.44\,\text{au}$ and $a_\text{out}=1.509a_\text{J}=7.85\,\text{au}$. We then simulate planetesimals with varying initial semi-major axes, randomly distributed across the range $a\in[0.6a_\text{J}, 1.55a_\text{J}]=[3.12\,\text{au}, 8.06\,\text{au}]$, assuming a surface density that follows the relationship $\displaystyle\Sigma\propto r^{-1}$ \citep{Bell+1997}. For each interval of $0.05a_\text{J}=0.26\,\text{au}$, $2,000$ planetesimals are introduced into the simulation, totalling $38,000$ planetesimals. Standard canonical units are used in our numerical simulations (for mass, distance, and time, normalizing the total system mass, the average Sun-Jupiter distance, and $\frac{1}{2\pi}$ of their orbital period to unity). The evolution of the system is simulated for $10^{6}T_\text{J}=11.86\,\text{Myr}$ using the $\textsc{Rebound}$ package (\citealt{Rein+2012}), with the $\textsc{IAS15}$ integrator. The time steps are automatically determined by the $\textsc{IAS15}$ algorithm. The simulation is halted if a planetesimal collides with either the planet or the star. Based on \citet{Zhou+2007} which examines the evolution of planetesimals with $e<0.5$, we incorporate the orbital element distribution of high-eccentricity planetesimals into our analysis, retain planetesimals on hyperbolic orbits until they reach a boundary with a radius of $100$ units (planetesimals outside this boundary with $e<1$ are also retained).

To ascertain the minimum distance between the planetesimal and the star, we collect the distance $r$ from the star to the planetesimal at each simulation step. If $r$ falls below the threshold of $0.05a_\text{p}$, we opt to collect the pericentre distance $q=a(1-e)$ instead of $r$ to improve accuracy.

\subsection{Limiting Distance of Tidal Fragmentation}

\indent\indent When a planetesimal approaches a star, it can experience fragmentation once it reaches the Roche limit, at which point the tidal force from the star surpasses the planetesimal's gravitational force. For a spherical, spinless, viscous fluid object, the Roche limit is given by $d_{\text{Roche}}=1.05\left(M_{\text{s}}/\rho\right)^{1/3}$, where $\rho$ is the bulk density of the object (\citealt{Sridhar+1992}). For small bodies where friction and material strength are dominant, their cohesion can prevent themselves from being tidally fragmented within this limit. According to the elastic-plastic continuum theory, the tidal fragmentation limit for small bodies where the material strength plays a dominant role, and the limiting distance $d_{\text{limit}}(R)$ for an object dominated by its own gravity are given by \citet{Holsapple+2008} and ZL20:
\begin{align}
    d_{\text{str}}&=\left(\frac{\sqrt{3}}{4\pi}\right)^{\frac{1}{3}}\left(\frac{5C}{4\pi GR^2\rho^2}+\frac{2\sin\phi}{\sqrt{3}(3 - \sin \phi)}\right)^{-\frac{1}{3}}\left(\frac{M_{\text{s}}}{\rho}\right)^{\frac{1}{3}}
    \label{dstr}
    \\
    d_{\text{limit}}(R)&=
    \begin{cases}
        d_{\text{Roche}}\quad R\gtrsim 1\text{km}
        \\d_{\text{str}}\quad R\lesssim 1\text{km}
    \end{cases}
\end{align}

Here, $R$ represents the radius of the object, while $C$ represents cohesion strength, and $\phi$ represents the friction angle. The tidal fragmentation limit depends on these parameters and decreases as the friction angle $\phi$ or the cohesion strength $C$ increases. This limit is smaller than the Roche limit in the presence of friction and cohesion.

Thermal modeling conducted by ZL20 indicates that the melting and re-condensation of surface silicates result in the formation of desiccated crusts on the fragments, thereby transforming them from cometary to asteroidal exteriors. This transition occurs about 3 hours after periastron, coinciding with an increase in cohesion strength from $C=0\,\text{~Pa}$ to $C=5.2\,\text{Pa}$. A parent body with $R=100\,\text{m}$, $\rho=2,000\,\mathrm{kg\,m^{-3}}$, $\phi=35^{\circ}$, $C=0\,\text{Pa}$ is efficient in producing `Oumuamua-like objects through tidal fragmentation process that includes the aforementioned thermal effects. Meanwhile, there is no sufficient evidence of whether the tidal fragmentation caused by a gas giant like Jupiter is adequate to produce elongated objects with comparable sizes and short-to-long axis ratios.

For most solar system bodies, $d_{\text{limit}}\lesssim R_{\odot}$, indicating that they would be squeezed onto the surface of the Sun before being torn apart by tidal force. This explains why we rarely observe small objects with extreme shapes like `Oumuamua in the Solar System. In contrast, objects around low-mass stars ($M_{\text{s}}<0.8M_\odot$) can be effectively torn apart by tidal force into sub-kilometre-sized fragments.

Tidal fragmentation can alter the energy of fragments, either increasing or decreasing it. However, simulation conducted by ZL20 shows that the change in orbital eccentricity due to tidal fragmentation is of very small magnitudes, on the order of $10^{-3}\sim10^{-6}$. Consequently, we disregard the impact of tidal fragmentation on orbital eccentricity when compared to the effect of giant planets, which can significantly excite the orbital eccentricities of planetesimals.

\subsection{Results of Sun-Jupiter System}

\begin{figure}
    \centering
    \includegraphics[width=0.45\textwidth]{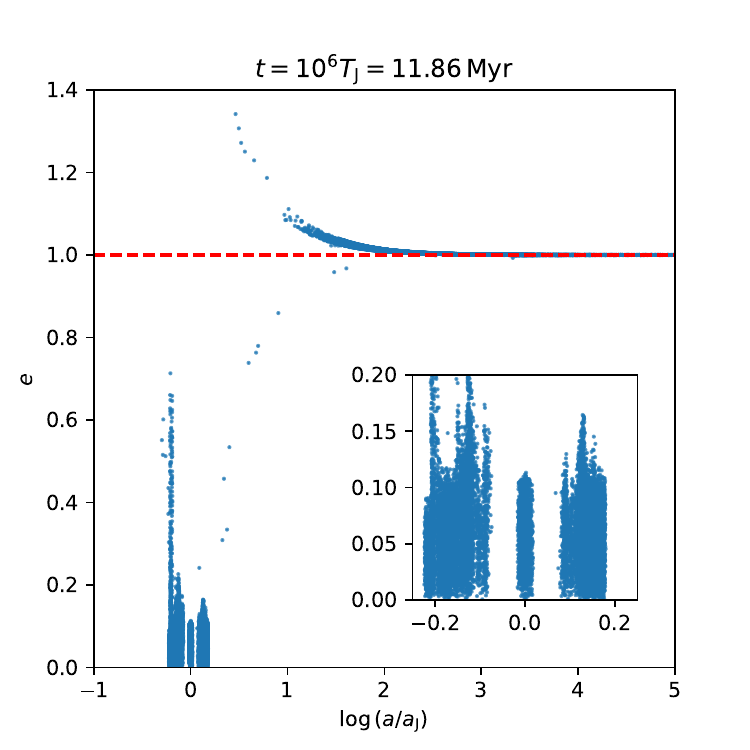}
    \caption{\label{2.1}Distribution of survived planetesimals' orbital elements $(a,e)$ in the Sun-Jupiter system at $t=10^{6}T_\text{J}$. The panel in the bottom right corner shows localized data points, which include resonant excitations.}
\end{figure}

\begin{figure}
    \centering
    \includegraphics[width=0.45\textwidth]{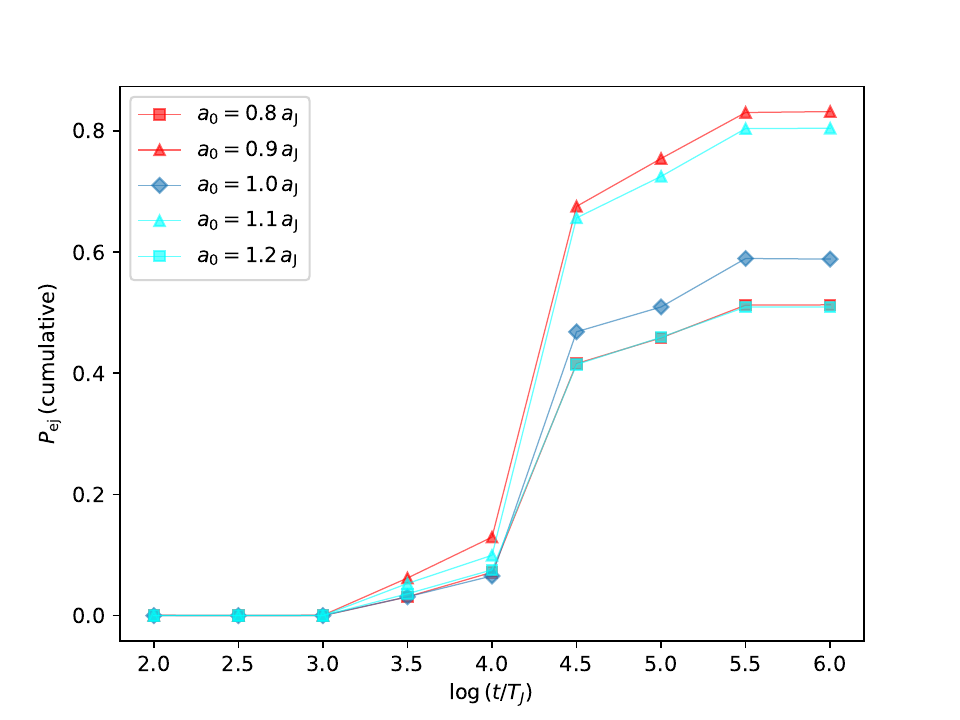}
    \includegraphics[width=0.45\textwidth]{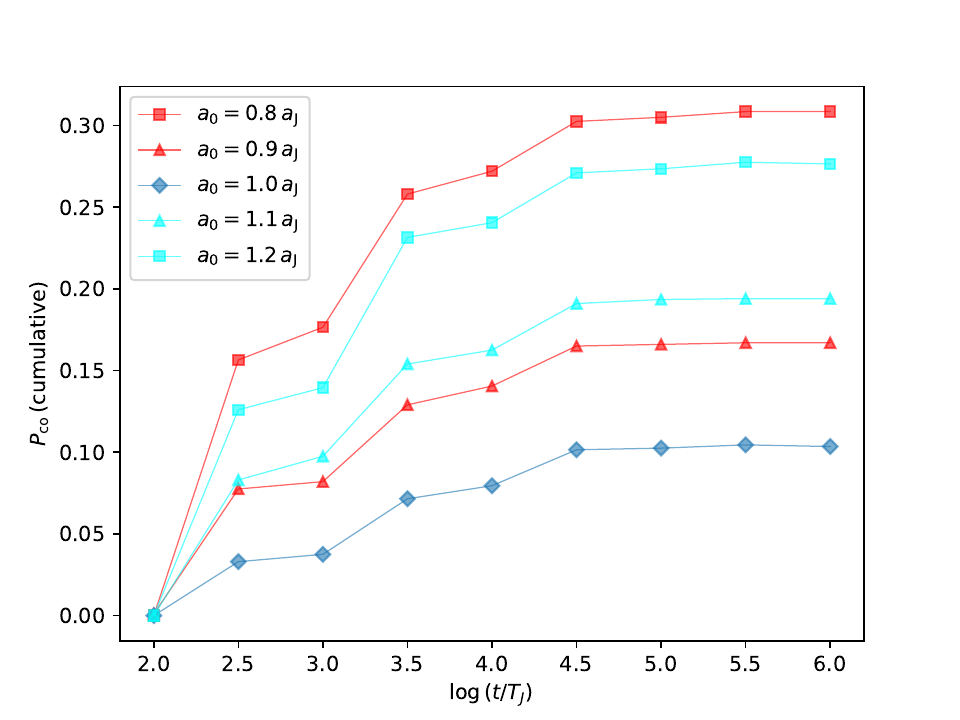}
    \caption{\label{2.3_2.4}The cumulative fraction of planetesimals. Upper panel: ejected; lower panel: collide in the Sun-Jupiter system at different periods. Here, the values of $a_0$ represent the range of $[a_0, a_0+0.5]$.} 
\end{figure}
\begin{figure}
    \centering
    \includegraphics[width=0.45\textwidth]{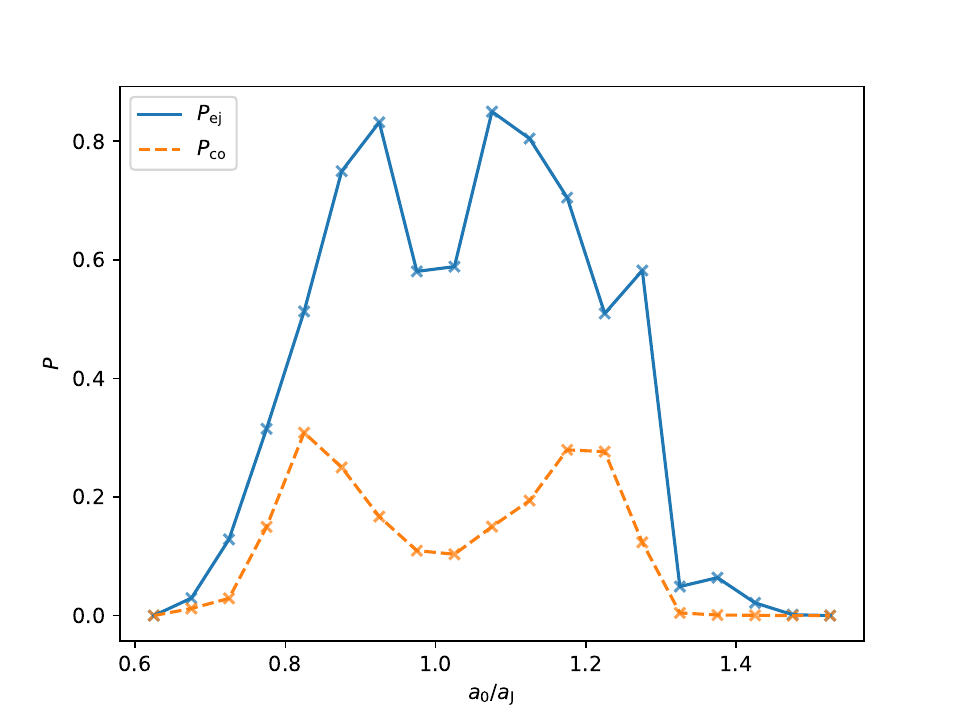}
    \caption{\label{2.5}The probabilities of collision and ejection of planetesimals with different initial semi-major axes determined at $10^{6}T_\text{J}$ in the Sun-Jupiter system. Yellow dashed line for collision probability, blue line for ejection probability.}

\end{figure}

\begin{figure}
    \centering
    \includegraphics[width=0.4\textwidth]{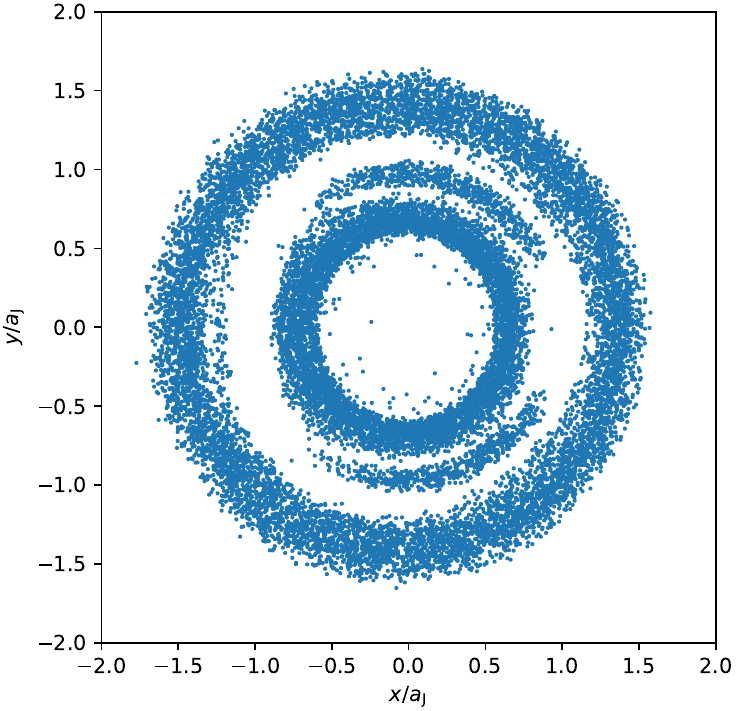}
    \caption{\label{2.7}The coordinates of the survived planetesimals in the rotating coordinate system after the evolution in the Sun-Jupiter system.}
\end{figure}

\indent\indent The simulation results (see Fig. \ref{2.1}) show that the orbital elements of planetesimals march on a specific rough routine on the $a$-$e$ plane. Most planetesimals over the line $e=1$ have either escaped the solar system on hyperbolic orbits or are in the process of escaping, and they can be classified as interstellar small objects. Given the long-term evolution we have simulated, the number of planetesimals on instantaneous hyperbolic orbits that will not evolve into interstellar minor bodies is negligible. The eccentricities of the planetesimals ejected from the solar system are statistically close to $1$. As for semi-major axis values, there is a significant range, from $10^1$ to $10^5$ times of Jovian orbital radius. Most of the survived planetesimals maintain $e<1.1$ and $a>10a_\text{J}$, which supports the hypothesis in \citet{Hallatt+2020} that most interstellar small objects have a low relative velocity at the moment of ejection. The proportion of planetesimals that were ejected and those colliding at different periods are presented in Fig. \ref{2.3_2.4}. Both the collision and ejection probabilities exhibit Gaussian-like distributions with respect to the initial semi-major axis but with a pronounced dip at $a_0/a_\text{p}=1.0$ in Fig. \ref{2.5}, at the same time, the challenge of exciting and colliding planetesimals out of the chaotic zone by Jupiter's gravitational diffusion is evidenced. We observe that near the planet, the probabilities of collision and ejection decrease, creating the valleys visible in Fig. \ref{2.5}. This could be attributed to $1:1$ resonance capture, as shown in Fig. \ref{2.7}.

The statistical outcomes of planetesimals with different evolutionary outcomes are detailed below: the proportion of 1) planetesimals that survived through evolution is $46.86\%$. The proportion of 2) planetesimals that collide with the planet is $11.87\%$, 3) collide with the star is $0.12\%$. 4) Ejected planetesimals constitute $41.15\%$. By collecting the distances or pericentres of planetesimals from the Sun every simulation step (since nondimensionalization conditions are used, the simulation results can be prolonged to other stellar systems), the probability of tidal fragmentation can then be assessed based on a comparison between the minimum distances and the tidal fragmentation limit of the star. Assuming the bulk density of planetesimal $\rho=2,000\,\mathrm{kg\,m^{-3}}$, the Roche limit of the Sun is $7.02\times10^{-3}\,\text{au}$. Considering internal cohesion and friction angle with $C=0\,\text{Pa}$ and $\phi=35^{\circ}$, the tidal fragmentation limit is calculated as $d_{\text{str}}=0.758d_{\text{Roche}}=5.32\times10^{-3}\,\text{au}$ (Eq. \ref{dstr}), that is, the consideration of friction does not alter the fragmentation limit's order of magnitude. No ejected planetesimal in the chaotic zone reaches the zone within $d_{\text{str}}$, thus, within a time-scale of approximately $10^{6}T_\text{J}=11.86\,\text{Myr}$, the proportion of 5) ejected planetesimals that experience tidal fragmentation is $0.00\%$.

\section{Evolution of Planetesimals in Exoplanet Systems}
\label{sect:3}

\subsection{System with a Low-Mass Star and a Hot Jupiter}
\label{sect:3.1}

\indent\indent In the simulation of the Sun-Jupiter system, we found that a fraction of planetesimals have minimum distances to the Sun that are less than $0.2a_\text{J}$. Based on this result, we can make a rough estimation: under a nondimensionalization scenario where the ratio of the stellar tidal fragmentation limit to the planetary orbital semi-major axis is taken as approximately $0.2a_\mathrm{p}$, then a fraction of the planetesimals in the chaotic zone may experience tidal fragmentation and being ejected. To meet the condition $0.2a_\text{p}\approx1.02\times10^{-3}a_\text{J}$, we set $a_\text{p}=0.005a_\text{J}=0.026\,\text{au}$. It is unlikely for hot Jupiters to have a significant number of planetesimals in their vicinity (\citealt{Zhou+2005}; \citealt{Raymond+2009}; \citealt{Raymond+2012}). However, for the sake of computational completeness, we still include hot Jupiters within our computational scope.

We focus on a system of a low-mass star ($M_{\text{s}}=0.5M_\odot$, $R_{\text{s}}=0.5R_\odot$), which is more likely to fragment a planetesimal without colliding with it, and a hot Jupiter ($M_\text{p}=M_\text{J}$, $R_\text{p}=R_\text{J}$). The semi-major axis of the hot Jupiter is set as $0.005a_\text{J}=0.026\,\text{au}$. From Eq. \ref{chaotic} we simulate semi-major axes within the range $a \in\left[0.5 a_{\text{p}}, 1.85 a_{\text{p}}\right]=\left[0.013\,\text{au}, 0.048\,\text{au}\right]$, with $54,000$ planetesimals put into the simulation.

\citet{Frewen+2014} and \citet{Rodet+2023} simulated systems of different planetary eccentricities, which shows the trend of increasing ejection rate with the increasing of planetary eccentricity at $0<e_\text{p}<0.2$, and the star-fragmentation rate increases with planetary eccentricity in the range $0<e_\text{p}<0.6$. Informed by their findings and the statistical result of exoplanets' median eccentricity (\citealt{Udry+2007}; \citealt{Winn+2015}), in the subsequent part of our study, we set the eccentricities of exoplanets as $0.2$.

\begin{figure}
    \centering
    \includegraphics[width=0.45\textwidth]{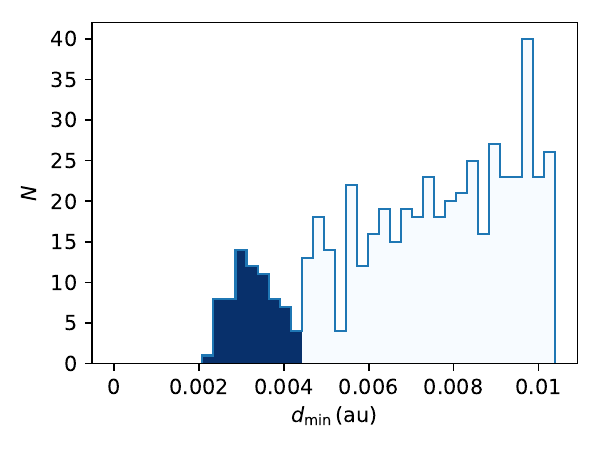}
	\caption{\label{4hothist}Statistics of planetesimals' minimum distances from the star in the system of a low-mass star and a hot Jupiter ($a_\text{p}=0.026\,\text{au}$), only planetesimals with eventual ejection are presented. The dark blue hists indicate that these planetesimals have moved within the star's tidal fragmentation limit during their evolutions.
}
\end{figure}

The proportion of 1) planetesimals that collide with the planet is as high as $68.09\%$, while 2) planetesimals that collide with the star is $3.65\%$. There are $12.91\%$ of the 3) planetesimals which survived through evolution. 4) Ejected planetesimals constitute $15.34\%$, and ejected planetesimals that experience tidal fragmentation account for $0.13\%$ of all planetesimals. These results suggest that collisions are the predominant mechanism in hot Jupiter systems. Fig. \ref{4hothist} shows there are 5) $0.13\%/15.34\%=0.83\%$ of the ejected planetesimals that experience $d_{\min}$ within the tidal fragmentation limit of the low-mass star ($4.22\times10^{-3}\,\text{au}$), leading to tidal fragmentation.

In hot Jupiter systems, collisions with hot Jupiters account for the majority of the planetesimals, thereby suppressing the scattering effect. Most planetesimals with small $d_{\min}$ experience early collisions with the giant planet and fail to achieve the expected $d_{\min}$ distribution.

\subsection{System with a Low-Mass Star and a Hot Neptune}

\indent\indent Compared to our initial simulation, we speculate that a system with a planet of radius smaller than Jupiter around a low-mass star would avoid the dominance of collisions and produce more ejected planetesimals. However, \citet{Safronov+1972} provided a parameter, the Safronov number $\Theta=\frac{a_\text{p}}{R_\text{p}} \frac{M_\text{p}}{M_{\text{s}}}$, to measure the ability of a planet to perturb surrounding bodies, or, in the context of our concern, to scatter small objects (\citealt{Ford+2008}; \citealt{Laughlin+2017}). From the perspective of the Safronov number, for planets on the same orbit, a smaller planetary mass corresponds to a smaller Safronov number and thus a weaker ability to scatter small objects.

To test this hypothesis, we attempt to simulate a hot Neptune system, checking whether it reduces the impact of collisions on high eccentricity planetesimals. We set the semi-major axis of the hot Neptune same as the hot Jupiter in the previous subsection. The radius and mass of the hot Neptune are taken as $R_\text{p}=0.3R_\text{J}$ and $M_\text{p}=0.08M_\text{J}$, respectively. The Safronov number of this hot Neptune is $0.0278$, while Safronov number of the hot Jupiter described in Section \ref{sect:3.1} is $0.106$. The chaotic zone of the system can be calculated using Eq. \ref{chaotic} to be $[0.609a_\text{p}, 1.572a_\text{p}]=[0.0158\,\text{au}, 0.0409\,\text{au}]$, and our simulation interval is set from $0.6a_\text{p}$ to $1.6a_\text{p}$ ($0.0156\,\text{au}$ to $0.0416\,\text{au}$).

The results presented that the proportion of 1) surviving planetesimals is $49.77\%$, while the proportion of 2) planetesimals colliding with the planet and 3) the star are $48.44\%$ and $0.48\%$, respectively. 4) Ejected planetesimals account for $1.31\%$, and 5) no ejected planetesimal experiences tidal fragmentation. In this case, interactions with the hot Neptune lead to collisions before most planetesimals can achieve hyperbolic orbits. Despite the smaller radius of a hot Neptune compared to a hot Jupiter, the slower excitation of eccentricity ($\sim1\%$ of a Jovian system according to the simulation results) still makes collisions the predominant outcome.

Therefore, even with a smaller radius, the slower excitation rate in a hot Neptune system maintains the prevalence of collisions, suggesting that the suppression of scattering by collisions is not solely determined by the planet's size but is also significantly influenced by the rate of eccentricity excitation. Despite the challenges a Neptune-mass planet faces in generating `Oumuamua-like objects through scattering, the significantly higher occurrence rate of close-in Neptunes or super-Earths compared to hot Jupiters means their contribution to interstellar small bodies cannot be neglected (\citealt{Zhu+2021}).

\subsection{System with a Low-Mass Star and a Warm Jupiter}

\indent\indent A warm Jupiter is typically defined as a gas giant with an orbital period $P$ between $10$ to $100$ days (\citealt{Jordan+2020}), which corresponds to a semi-major axis $a$ ranging from $0.02a_\text{J}$ to $0.1a_\text{J}$. In our simulation, we set $a_\text{p}=0.1a_\text{J}=0.52\,\text{au}$ and assume the mass and radius of the warm Jupiter are the same as Jupiter, while the parameters of the low-mass star remain consistent with those previously stated ($M_{\text{s}}=0.5M_\odot$, $R_{\text{s}}=0.5R_\odot$). The Safronov number of the warm Jupiter is calculated to be $2.12$, indicating a stronger scattering ability compared to aforementioned close-in planets.

\begin{figure}
    \centering
    \includegraphics[width=0.5\textwidth]{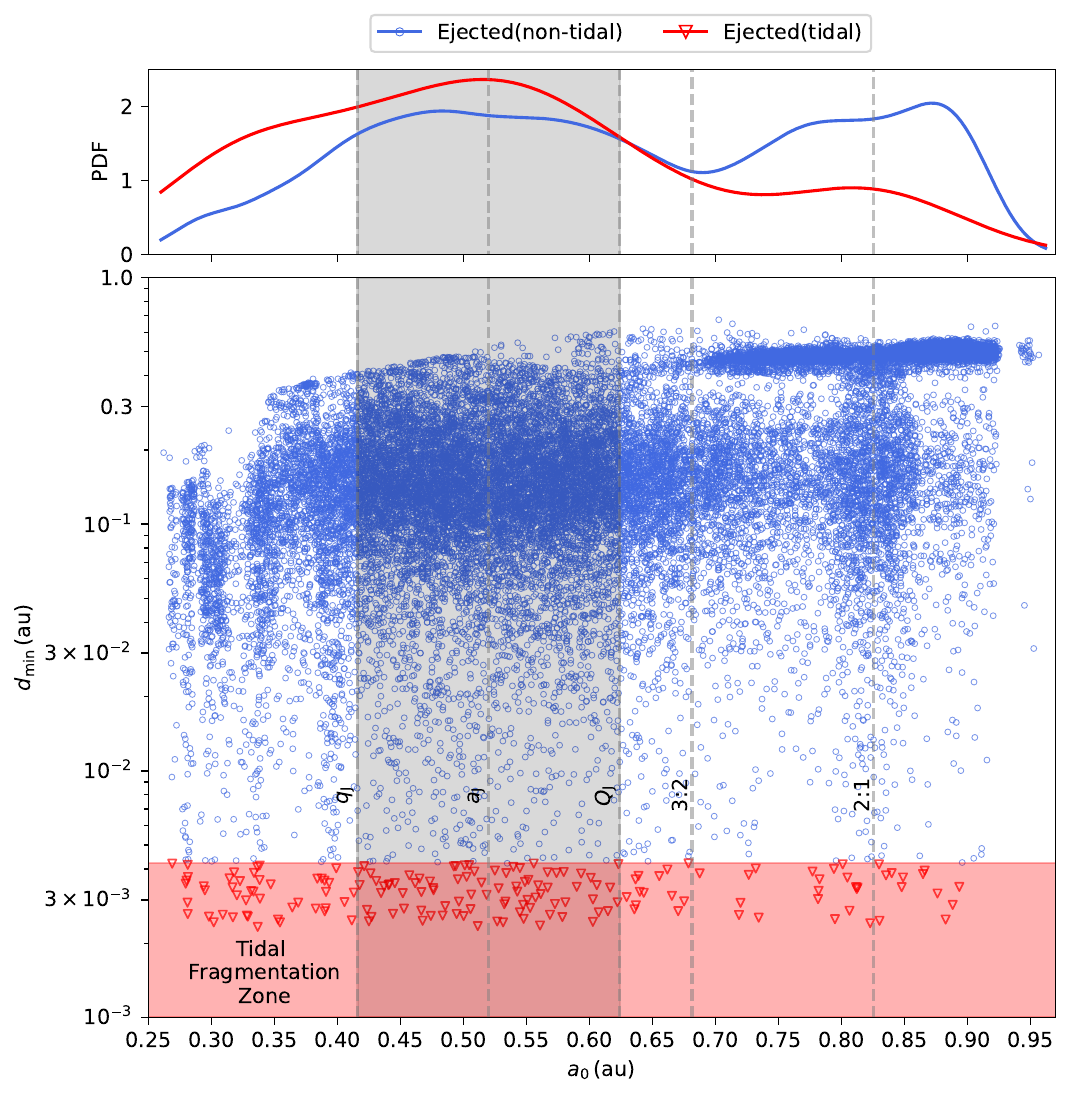}
	\caption{\label{4warm}Minimum distances of the ejected planetesimals from the star in the system of a low-mass star and a warm Jupiter, with different initial semi-major axes of planetesimals. The perihelion distance, semi-major axis, aphelion distance of the warm Jupiter, as well as the locations of the $2:1$ and $3:2$ resonances with the warm Jupiter, are all indicated by gray dashed lines.}
\end{figure}

The simulation results are as follows: the proportion of 1) survived planetesimals is 13.02\%, while the proportion of 2) planetesimals colliding with the planet and 3) the star are 28.65\% and 5.96\%, respectively. 4) Ejected planetesimals account for 52.37\%. Fig. \ref{4warm} illustrates the minimum distances of planetesimals from the star. The boundary outlined by the planetesimals experiencing tidal fragmentation in the figure represents the tidal fragmentation limit. Reading the number of planetesimals below the tidal fragmentation limit in Fig. \ref{4warm}, there are 5) $0.62\%$ of the ejected planetesimals that undergo tidal fragmentation (to the number of ejected planetesimals).

\subsection{Error Analysis and Other System Configurations}

\begin{figure*}
    \centering
    \includegraphics[width=0.8\textwidth]{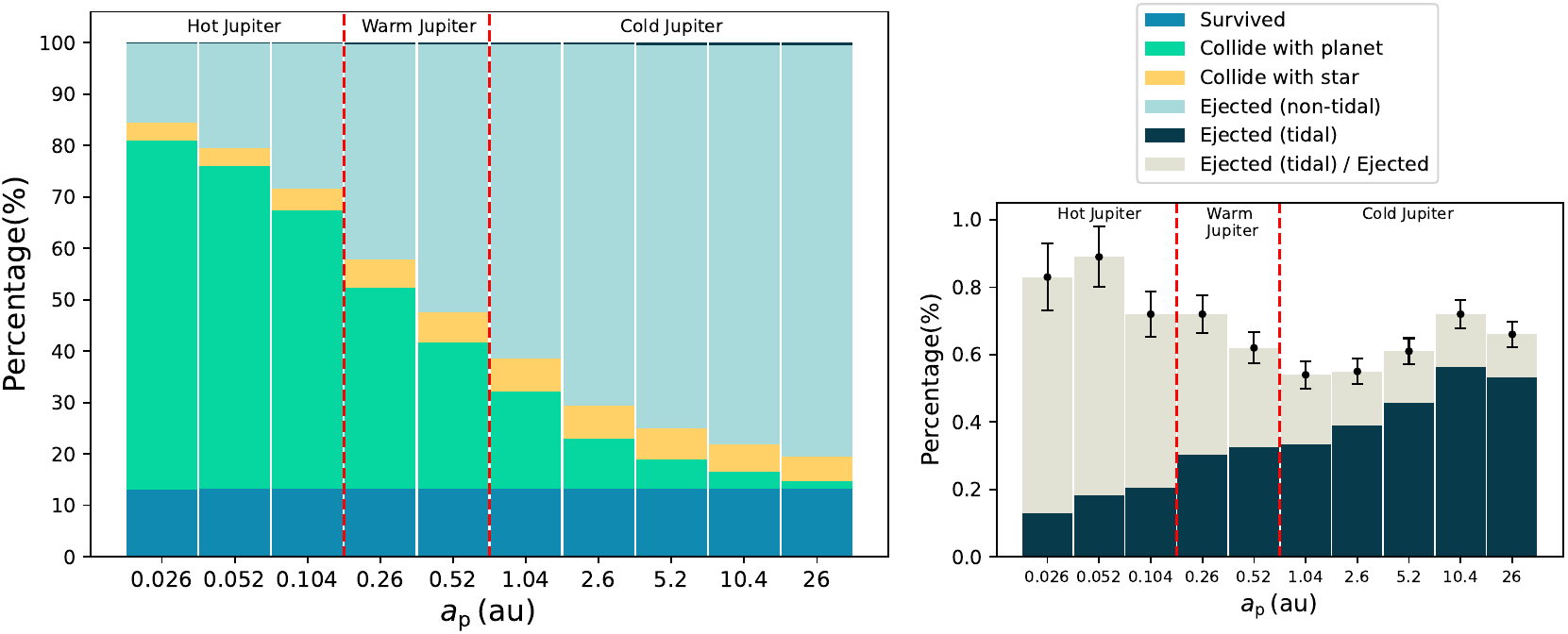}
	\caption{\label{4all}The results (the probability of a planetesimal in the chaotic zone evolved to be ejected, and the probability of an ejected planetesimal undergoing tidal fragmentation, the probability of colliding with the planet, with the star and surviving) of a system with a low mass star and an eccentric Jupiter with different planetary semi-major axes. The lower right panel emphasizes the proportion of ejected planetesimals that have experienced tidal fragmentation, with the yellow bars indicating the percentage of ejected planetesimals that have been tidally fragmented out of all ejected planetesimals.}
\end{figure*}

\begin{figure*}
    \centering
    \includegraphics[width=0.95\textwidth]{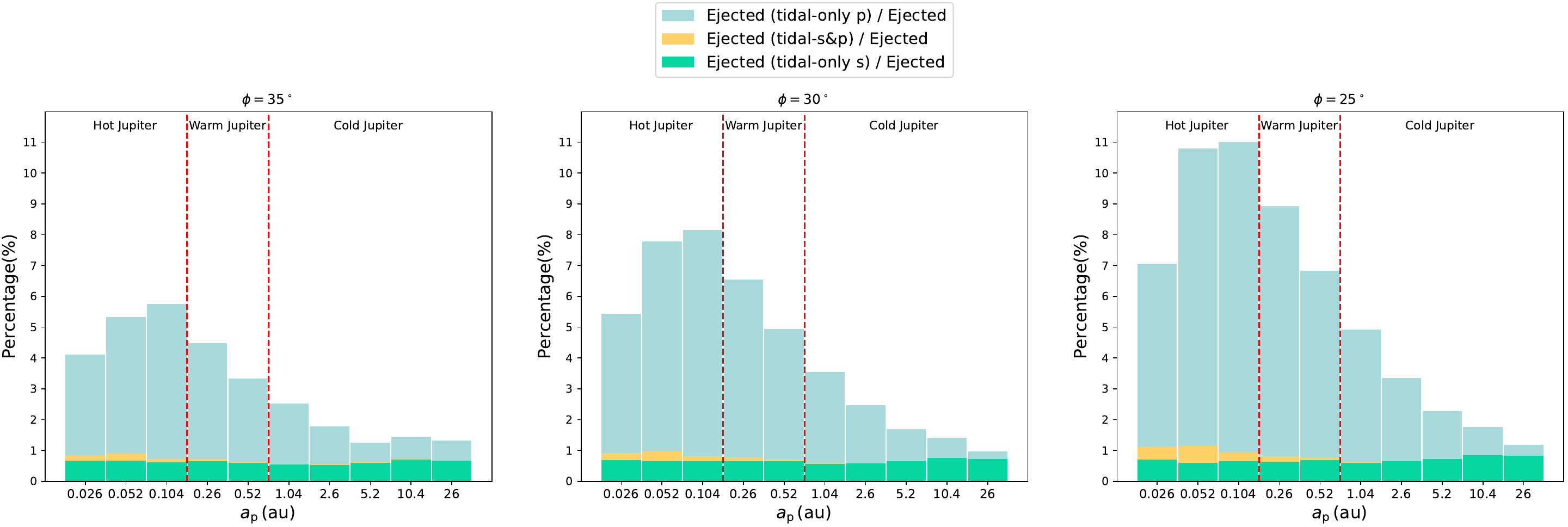}
	\caption{\label{4add}The probability of an ejected planetesimal undergoing different types: only by the star (tidal-only s), only by the planet (tidal-only p), and both (tidal-s\&p), of tidal fragmentation. The left and right panels represent the results corresponding to different internal friction angles of planetesimals. It should be noted that these probabilities (including those shown in the figure) do not overlap with each other.
}
\end{figure*}

\indent\indent In addition to the systems above, we continued to simulate hot, warm, and cold Jupiter systems with varying orbital semi-major axes. The results are presented in Fig. \ref{4all} and Table \ref{table1}. The standard deviation and the corresponding confidence interval of each possibility can be estimated as $\sigma=\sqrt{\frac{p(1-p)}{n}}$ and $1.96\sigma$. The number of simulated planetesimals in a Jovian system is $n=54,000$ as mentioned above. When considering the proportion of ejected planetesimals that are subject to tidal fragmentation, we replace $n$ with the number of ejected planetesimals $P(\text{ej}) n$ since the number of planetesimals ejected is less than the total number of planetesimals. 

An example planetesimal in the system of $a_\text{p}=2a_\text{J}=10.4\,\text{au}$, $e_\text{p}=0.2$ that undergoes eccentricity excitation, tidal fragmentation, and eventually gets ejected from the system is shown in Fig. \ref{example}. Considering that the probability of an ejected planetesimal undergoing tidal fragmentation $P(\text{tidal}|\text{ej})$ reaches its local maximum value at the condition of $a_\text{p}=2a_\text{J}=10.4\,\text{au}$, $e_\text{p}=0.2$, we conducted additional simulations with different planetary masses ($2M_\text{J}$ and $4M_\text{J}$) based on $a_\text{p}=2a_\text{J}$. Due to the different boundaries of chaotic zones corresponding to systems with different planetary masses, we randomly place $50,000$ planetesimals within the chaotic zone of each system based on the semi-major axes. The results are shown in Table \ref{table2}.

\subsection{Different Material Properties and Tidal Fragmentation by the Planet}

\indent\indent Although we have examples of tidal fragmentation by a planet like the comet P/Shoemaker-Levy 9 (\citealt{Weaver+1995}), the debris cannot maintain stable elongated shapes. Whether planets can produce `Oumuamua-like objects through tidal fragmentation is currently lacking in both simulation and observational evidence.

According to Eq. \ref{dstr}, under the assumptions of our model, for Jupiter to produce a similar tidal fragmentation process (without considering the temperature difference between Jupiter and a star), its tidal fragmentation limit is $5.20\times10^{-4}\,\text{au}$, while its own radius is $4.77\times10^{-4}\,\text{au}$. That is, it has the potential to fragment planetesimals. Therefore, in our simulations, we have included the collection of the distance between the planetesimals and the planet (collecting at each simulation step, and when the distance is less than $0.005a_\text{p}$, we collect the osculating perihelion distance of the planetesimal relative to the planet's orbit). We then analyse whether each planetesimal is fragmented by the star or by the planet under conditions where the friction angles $\phi = 35^\circ$, $30^\circ$ and $25^\circ$. The results are shown in Fig. \ref{4add}, Table \ref{table3}, Table \ref{table4} and Table \ref{table5}.

\section{Discussion}
\label{sect:4}

\indent\indent Throughout the evolution we presented in Section \ref{sect:2}, collision events distributed approximately uniformly over time. However, ejection events peak around $10^4\sim10^5T_\text{J}=0.1\sim1\,\text{Myr}$ (Fig. \ref{2.3_2.4}). For eccentric Jupiter systems, We plot similar diagrams and find that the range of time-scale is $10^3\sim10^{4.5}T_\text{p}$. These findings provide valuable insights into our analysis of the age of interstellar objects such as 1I/`Oumuamua. Assuming a similar cold Jupiter system as its source, the age of an interstellar object would be the time-scale of the sum of the ejection process and the travelling process between stellar systems.

We have calculated that in low-mass stellar systems, a single Jupiter-like planet can eject a portion of the planetesimals in its surrounding protoplanetary disc beyond the system, thereby converting them into interstellar small bodies. A fraction of these planetesimals undergoes tidal fragmentation by the host star, with the fraction evaluated by (only considering warm Jupiter and cold Jupiter systems):
\begin{align}
    P(\text{tidal}|\text{ej})=\left(\frac{\phi}{35^\circ}\right)^{-0.50\pm0.06}\left(\frac{M_\text{p}}{M_\text{J}}\right)^{-0.69\pm0.29}0.63\%\pm0.14\%\label{ptidal}
\end{align}

The values of $\phi$ and $M_\text{p}$ are both within the range limited by the simulations in this paper ($\phi\in\left[25^\circ,35^\circ\right]$ and $M_\text{p}\in[M_\text{J},4M_\text{J}]$). The fraction $0.63\pm0.14\%$ is based on the results in Fig. \ref{4all} and Table \ref{table1}, and the relationship between the probability and $\phi$, $M_\text{p}$ is estimated from Fig. \ref{4add}, Table \ref{table2}, Table \ref{table3} and Table \ref{table4}. The correlation between this ratio and the orbital semi-major axis is weak, reaching its maximum at planetary mass $\sim M_\text{J}$. We used the $\chi^2$ test to obtain a $p$-value of $2\times10^{-4}$, indicating that for warm Jupiters and cold Jupiters with different semi-major axes, we can consider $P(\text{tidal}|\text{ej})$ a constant (this may not apply to long-period cold Jupiters with semi-major axes $a_\text{p}>5a_\text{J}=26\,\mathrm{au}$). The proportion of survived planetesimals also remains nearly constant: during the evolution, the likelihood of close encounters between planetesimals and the two primary bodies is almost independent of the planetary semi-major axis, after being disturbed during close encounters, planetesimals are either scattered or collide with two primary bodies, remaining survived planetesimals of similar proportions. When other conditions are held constant, $P(\text{ej})$ and $\log(a_\text{p}/a_\text{J})$ exhibit an approximate Logistic relationshop. This relationship also holds for colliding planetesimals, reflecting that under non-dimensionalized conditions, as the system scale we take becomes larger, the non-dimensionalized radius of the primary bodies becomes smaller, the probability of planetesimals colliding with the primary bodies becomes lower, and the ejection probability correspondingly becomes higher.

Suppose each fragmented planetesimal generates $N$ fragments, with $N_\text{e}$ of these exhibiting extreme axis ratios ($\gtrsim5:1$). The proportion of `Oumuamua-like fragments with extreme axis ratios among all the small bodies ejected beyond the system is:
\begin{align}
P_{\text{e}}=\frac{P(\text{tidal}|\text{ej})N_\text{e}}{1-P(\text{tidal}|\text{ej})+P(\text{tidal}|\text{ej})N}\approx\frac{P(\text{tidal}|\text{ej})N_\text{e}}{1+P(\text{tidal}|\text{ej})N}
\label{pe}
\end{align}

Observational and theoretical studies have revealed that the outcomes of tidal fragmentation events exhibit significant variability based on several key parameters: size, density, the specifics of the encounter trajectory (\citealt{Asphaug+1996}; \citealt{Walsh+2018}). The rotation state and the orientation of the object's principal axis at the periastron significantly impact mass loss patterns: prograde rotation enhances disruption potential, and retrograde rotation provides a stabilizing effect (\citealt{Richardson+1998}). According to the simulation results in ZL20, parent bodies with radii between $R=100\,\text{m}$ to $R=10\,\text{km}$ can produce `Oumuamua-like fragments by order of magnitude of $N_\text{e}\sim 10^1$, while $N\sim10^2$. By substituting the moderate values of parameters $\phi=30^\circ$ and $M_\text{p}=2M_\text{J}$, we estimate that the proportion of `Oumuamua-like fragments among all interstellar small bodies generated by a system with a low-mass star and a Jovian planet is $\sim3\%$.

Both $N$ and $N_\text{e}$ are functions of $\phi$. Currently, we do not have sufficient information to specify the exact relationship between $N$, $N_\text{e}$, and $\phi$. This relationship can be refined as more simulation results become available.

The probability of planetesimals being tidally fragmented by a planet is related to the semi-major axis of the planet's orbit. Considering the cases of warm and cold Jupiters, the relationship between probability and semi-major axis can be estimated as $P(\text{tidal-p}|\text{ej})\propto a_{\text{p}}^{-0.52\pm0.06}$. The coefficient of determination $R^2$ values for this fitting relationshop, including the one in Eq. \ref{ptidal}, are all greater than $0.995$. Suppose it is confirmed that tidal fragmentation of planets can produce similar elongated fragments. In that case, we can use the data from Fig. \ref{4add}, Table \ref{table3}, and Table \ref{table4} to calculate the proportion of such fragments relative to total fragments. For different friction angles, the probability of an ejected planetesimal undergoing tidal fragmentation by a warm Jupiter is significantly higher than that by the star in the system ($3\%\sim8\%$). For cold Jupiters, when the planetary semi-major axis is less than $10\,\text{au}$, the events of tidal fragmentation are also dominated by the planet rather than the star. The majority of exoplanets detected by current observational methods have semi-major axes within this threshold (\citealt{Zhu+2021}). This means that for the currently known planetary systems, tidal fragmentation by planets is more frequent.

For comparison, combining a large number of simulations, \citet{Raymond+2018} showed that $0.1\%\sim1\%$ of cometary planetesimals undergo disruptive encounters before ejection. The differences in system setup between \citet{Raymond+2018} and this work, apart from the star and planets, the set of outer discs of planetesimals also have a noticeable impact on the results: the probability density function plotted in Fig. \ref{4warm} shows a lower probability of the outer disc being fragmented by the star. The tidal fragmentation limit distance calculated (with bulk density of planetesimals $1,000\,\mathrm{kg\,m^{-3}}$, no friction) is greater than we used in this work. These factors combined may account for the differences in the probabilities calculated by the two works.

Incorporate planetary tidal fragmentation into consideration, the probability calculated by Eq. \ref{pe} correspondingly becomes:

\begin{align}
P_{\text{e}}=\frac{P(\text{tidal-s}|\text{ej})N_\text{e-s}+P(\text{tidal-p}|\text{ej})N_\text{e-p}}{1+P(\text{tidal-s}|\text{ej})N_{\text{s}}+P(\text{tidal-p}|\text{ej})N_{\text{p}}}
\end{align}

The subscripts $s$ and $p$ represent the effects of the star and the planet, respectively. Due to the lack of evidence for the formation of extremely elongated objects by planetary tidal fragmentation, we take $N_\text{e-p}=0$ and $N_{\text{p}}\sim10^{-1}N_{\text{s}}$. For cold Jupiters, $P(\text{tidal-p}|\text{ej})\sim P(\text{tidal-s}|\text{ej})$, tidal fragmentation by the planet does not significantly affect the value of $P_\text{e}$. However, for warm Jupiters, $P(\text{tidal-p}|\text{ej})\sim 10P(\text{tidal-s}|\text{ej})$, which reduces $P_{\text{e}}$ by $\sim30\%$ of its original value.

\section{Conclusions and Implications}
\label{sect:5}

\indent\indent By analyzing the scattering process of a giant planet, we have obtained statistical information regarding the formation of interstellar small objects, which has the potential to help us determine their ages. With existing studies of tidal fragmentation, we have made attempts to evaluate the possibility of this mechanism, then determine the proper condition for generating `Oumuamua-like elongated objects. Through simulations of single giant planet systems with different configurations, we conclude that if the parent body of an elongated small object like `Oumuamua originates from a planetesimal disc around a single giant planet, the likelihood of the giant planet being an eccentric Jupiter ($e\sim0.2$, $M_\text{p}\sim M_\text{J}$) is higher than other configurations. Some planetesimals are torn apart by the giant planet, producing more fragments (not elongated). According to ZL20, most fragments produced through tidal fragmentation exhibit ordinary axis ratios, with only a small portion torn into extremely elongated shapes. Adding various dissipative effects into account, the estimated proportion we provide is an upper limit. The plausibility of the event that the first interstellar object observed
being extremely elongated is weakly proved according to our result $P_\text{e}\sim3\%$. The relatively low probability we have derived in this work prompts us to further investigate different parameters, multi-planet systems and other mechanisms like close stellar flybys in clusters (\citealt{Pfalzner+2021}; \citealt{Jewitt+2023}). The parent body has other possible identities suggested by ZL20, including an object from the Oort cloud or a planet-sized super-Earth. Additionally, planetesimals in multi-planet or multi-star systems are also potential origins and are more efficient at in producing interstellar objects (\citealt{Jackson+2018}; \citealt{Levine+2023}).

For Solar-mass stars, we adopt the gas giant occurrence rate of $10\sim20\%$ from planet surveys (\citealt{Winn+2015}; \citealt{Fernandes+2019}; \citealt{He+2020}), and lower occurrence rate $<1\%$ of gas giants for M dwarfs (\citealt{Johnson+2010}; \citealt{Dressing+2015}; \citealt{Mulders+2015}; \citealt{Quirrenbach+2022}). The ratio of the number of G-type stars to M dwarfs ($\sim1:10$) is close to the ratio of the occurrence rate of giant planets in M dwarf systems to that in G-type star systems, which means that among these systems capable of generating interstellar planetesimals through planetary perturbations, the number of systems that can produce `Oumuamua-like objects through stellar tidal fragmentation is comparable to the number of systems that can only produce planetary tidal fragmentation processes. This analysis suggests that both mechanisms contribute meaningfully to the population of fragmented interstellar objects.

The relative contribution of these mechanisms also has implications for the physical properties of interstellar objects. Fragments generated through planetary fragmentation would retain compositional signatures of the outer planetary system regions where they originated, while stellar fragmentation fragments may sample a broader range of characteristics. Future characterization of interstellar objects may help distinguish between these formation channels.

Previous efforts have been made to search for possible origins of `Oumuamua using the orbit inversion method (\citealt{Zuluaga+2018}; \citealt{Feng+2018}; \citealt{Bailer+2018}; \citealt{Zwart+2018}). \citet{Hallatt+2020} incorporated factors such as the Galactic potential and disc heating in their simulations, identifying 70 potential progenitors. By comparing these sources with data from the NASA Exoplanet Archive, it was found that only one host star (HR 8799) is known with four cold Jupiters orbiting, while the planets in other systems are either nonexistent or undetected. A comprehensive understanding of the planets in these systems would enable further dynamical analyses of each system. Since the strength of the tidal fragmentation increases with the density of the host star, tidal fragmentation of `Oumuamua-like objects can only occur near stars with higher densities. Therefore, the potential progenitors identified through orbital inversion should be further filtered based on the parameters of the host star, focusing on candidates with higher densities, i.e., the most common M dwarf or K-type stars (\citealt{Chabrier+2003}). After this filtering and a more comprehensive understanding of these systems, future numerical simulations can provide better insights for assessing the feasibility of each system as the origin of `Oumuamua.

\section*{Acknowledgements}

We thank the referee, Sean N. Raymond, whose comments greatly improved the quality and clarity of the manuscript. This work was funded by the National Key R\&D Program of China under No.SQ2024YFA1600043, and the National Natural Science Foundation of China (NSFC) under No.12150009, 11933001, 11973028, and 1803012. We also acknowledge the science research grants from the China Manned Space Project with No. CMS-CSST-2021-B12, as well as Civil Aerospace Technology Research Project (D010102). We made use of the Python library $\textsc{NumPy}$ (\citealt{Harris+2020}), and the figures were made with $\textsc{Matplotlib}$ (\citealt{Hunter+2007}).

\section*{Data Availability}

The simulation files and results underlying this article have been uploaded on GitHub and can be found at \href{https://github.com/Xiling-Zheng/Zheng_and_Zhou2025.git}{$\mathtt{https://github.com/XilingZheng/ZhengAndZhouMNRAS2025.git}$}.



\bibliographystyle{mnras}
\bibliography{example} 

\begin{thebibliography}{}
\makeatletter
\relax
\def\mn@urlcharsother{\let\do\@makeother \do\$\do\&\do\#\do\^\do\_\do\%\do\~}
\def\mn@doi{\begingroup\mn@urlcharsother \@ifnextchar [ {\mn@doi@}
  {\mn@doi@[]}}
\def\mn@doi@[#1]#2{\def\@tempa{#1}\ifx\@tempa\@empty \href
  {http://dx.doi.org/#2} {doi:#2}\else \href {http://dx.doi.org/#2} {#1}\fi
  \endgroup}
\def\mn@eprint#1#2{\mn@eprint@#1:#2::\@nil}
\def\mn@eprint@arXiv#1{\href {http://arxiv.org/abs/#1} {{\tt arXiv:#1}}}
\def\mn@eprint@dblp#1{\href {http://dblp.uni-trier.de/rec/bibtex/#1.xml}
  {dblp:#1}}
\def\mn@eprint@#1:#2:#3:#4\@nil{\def\@tempa {#1}\def\@tempb {#2}\def\@tempc
  {#3}\ifx \@tempc \@empty \let \@tempc \@tempb \let \@tempb \@tempa \fi \ifx
  \@tempb \@empty \def\@tempb {arXiv}\fi \@ifundefined
  {mn@eprint@\@tempb}{\@tempb:\@tempc}{\expandafter \expandafter \csname
  mn@eprint@\@tempb\endcsname \expandafter{\@tempc}}}

\bibitem[\protect\citeauthoryear{Asphaug \& Benz}{Asphaug \&
  Benz}{1996}]{Asphaug+1996}
Asphaug E.,  Benz W.,  1996, \mn@doi [Icarus]
  {https://doi.org/10.1006/icar.1996.0083}, 121, 225

\bibitem[\protect\citeauthoryear{Bailer-Jones, Farnocchia, Meech, Brasser,
  Micheli, Chakrabarti, Buie  \& Hainaut}{Bailer-Jones
  et~al.}{2018}]{Bailer+2018}
Bailer-Jones C. A.~L.,  Farnocchia D.,  Meech K.~J.,  Brasser R.,  Micheli M.,
  Chakrabarti S.,  Buie M.~W.,   Hainaut O.~R.,  2018, \mn@doi [AJ]
  {10.3847/1538-3881/aae3eb}, 156, 205

\bibitem[\protect\citeauthoryear{Bannister et~al.,}{Bannister
  et~al.}{2017}]{Bannister+2017}
Bannister M.~T.,  et~al., 2017, \mn@doi [ApJL]
  {https://doi.org/10.3847/2041-8213/aaa07c}, 851, L38

\bibitem[\protect\citeauthoryear{Bell, Cassen, Klahr  \& Henning}{Bell
  et~al.}{1997}]{Bell+1997}
Bell K.~R.,  Cassen P.~M.,  Klahr H.~H.,   Henning T.,  1997, \mn@doi [ApJ]
  {10.1086/304514}, 486, 372

\bibitem[\protect\citeauthoryear{Bolin et~al.,}{Bolin
  et~al.}{2018}]{Bolin+2018}
Bolin B.~T.,  et~al., 2018, \mn@doi [ApJL] {10.3847/2041-8213/aaa0c9}, 852, L2

\bibitem[\protect\citeauthoryear{Chabrier}{Chabrier}{2003}]{Chabrier+2003}
Chabrier G.,  2003, \mn@doi [ApJ] {10.1086/374879}, 586, L133

\bibitem[\protect\citeauthoryear{Drahus, Guzik, Waniak, Handzlik, Kurowski  \&
  Xu}{Drahus et~al.}{2018}]{Drahus+2018}
Drahus M.,  Guzik P.,  Waniak W.,  Handzlik B.,  Kurowski S.,   Xu S.,  2018,
  \mn@doi [Nat Astron] {https://doi.org/10.1038/s41550-018-0440-1}, 2, 407

\bibitem[\protect\citeauthoryear{Dressing \& Charbonneau}{Dressing \&
  Charbonneau}{2015}]{Dressing+2015}
Dressing C.~D.,  Charbonneau D.,  2015, \mn@doi [ApJ]
  {10.1088/0004-637X/807/1/45}, 807, 45

\bibitem[\protect\citeauthoryear{Feng \& Jones}{Feng \&
  Jones}{2018}]{Feng+2018}
Feng F.,  Jones H. R.~A.,  2018, \mn@doi [ApJL] {10.3847/2041-8213/aaa404},
  852, L27

\bibitem[\protect\citeauthoryear{Fernandes, Mulders, Pascucci, Mordasini  \&
  Emsenhuber}{Fernandes et~al.}{2019}]{Fernandes+2019}
Fernandes R.~B.,  Mulders G.~D.,  Pascucci I.,  Mordasini C.,   Emsenhuber A.,
  2019, \mn@doi [ApJ] {10.3847/1538-4357/ab0300}, 874, 81

\bibitem[\protect\citeauthoryear{Fitzsimmons et~al.,}{Fitzsimmons
  et~al.}{2018}]{Fitzsimmons+2018}
Fitzsimmons A.,  et~al., 2018, \mn@doi [Nat Astron]
  {https://doi.org/10.1038/s41550-017-0361-4}, 2, 133

\bibitem[\protect\citeauthoryear{Ford \& Rasio}{Ford \&
  Rasio}{2008}]{Ford+2008}
Ford E.~B.,  Rasio F.~A.,  2008, \mn@doi [ApJ] {10.1086/590926}, 686, 621

\bibitem[\protect\citeauthoryear{Frewen \& Hansen}{Frewen \&
  Hansen}{2014}]{Frewen+2014}
Frewen S. F.~N.,  Hansen B. M.~S.,  2014, \mn@doi [MNRAS]
  {10.1093/mnras/stu097}, 439, 2442

\bibitem[\protect\citeauthoryear{Hallatt \& Wiegert}{Hallatt \&
  Wiegert}{2020}]{Hallatt+2020}
Hallatt T.,  Wiegert P.,  2020, \mn@doi [AJ] {10.3847/1538-3881/ab7336}, 159,
  147

\bibitem[\protect\citeauthoryear{Harris et~al.,}{Harris
  et~al.}{2020}]{Harris+2020}
Harris C.~R.,  et~al., 2020, \mn@doi [Nature]
  {https://doi.org/10.1038/s41586-020-2649-2}, 585, 357

\bibitem[\protect\citeauthoryear{He, Ford  \& Ragozzine}{He
  et~al.}{2020}]{He+2020}
He M.~Y.,  Ford E.~B.,   Ragozzine D.,  2020, \mn@doi [AJ]
  {10.3847/1538-3881/abc68b}, 161, 16

\bibitem[\protect\citeauthoryear{Holsapple \& Michel}{Holsapple \&
  Michel}{2008}]{Holsapple+2008}
Holsapple K.~A.,  Michel P.,  2008, \mn@doi [Icarus]
  {https://doi.org/10.1016/j.icarus.2007.09.011}, 193, 283

\bibitem[\protect\citeauthoryear{Hunter}{Hunter}{2007}]{Hunter+2007}
Hunter J.~D.,  2007, \mn@doi [Computing in Science \& Engineering]
  {10.1109/MCSE.2007.55}, 9, 90

\bibitem[\protect\citeauthoryear{Jackson, Tamayo, Hammond, Ali-Dib  \&
  Rein}{Jackson et~al.}{2018}]{Jackson+2018}
Jackson A.~P.,  Tamayo D.,  Hammond N.,  Ali-Dib M.,   Rein H.,  2018, \mn@doi
  [MNRASL] {https://doi.org/10.1093/mnrasl/sly033}, 478, L49

\bibitem[\protect\citeauthoryear{Jewitt \& Seligman}{Jewitt \&
  Seligman}{2023}]{Jewitt+2023}
Jewitt D.,  Seligman D.~Z.,  2023, \mn@doi [ARA\&A]
  {10.1146/annurev-astro-071221-054221}, 61, 197

\bibitem[\protect\citeauthoryear{Jewitt, Luu, Rajagopal, Kotulla, Ridgway, Liu
  \& Augusteijn}{Jewitt et~al.}{2017}]{Jewitt+2017}
Jewitt D.,  Luu J.,  Rajagopal J.,  Kotulla R.,  Ridgway S.,  Liu W.,
  Augusteijn T.,  2017, \mn@doi [ApJL] {10.3847/2041-8213/aa9b2f}, 850, L36

\bibitem[\protect\citeauthoryear{Johnson, Aller, Howard  \& Crepp}{Johnson
  et~al.}{2010}]{Johnson+2010}
Johnson J.~A.,  Aller K.~M.,  Howard A.~W.,   Crepp J.~R.,  2010, \mn@doi
  [PASP] {10.1086/655775}, 122, 905

\bibitem[\protect\citeauthoryear{Jordán et~al.,}{Jordán
  et~al.}{2020}]{Jordan+2020}
Jordán A.,  et~al., 2020, \mn@doi [AJ] {10.3847/1538-3881/ab6f67}, 159, 145

\bibitem[\protect\citeauthoryear{Knight, Protopapa, Kelley, Farnham, Bauer,
  Bodewits, Feaga  \& Sunshine}{Knight et~al.}{2017}]{Knight+2017}
Knight M.~M.,  Protopapa S.,  Kelley M. S.~P.,  Farnham T.~L.,  Bauer J.~M.,
  Bodewits D.,  Feaga L.~M.,   Sunshine J.~M.,  2017, \mn@doi [ApJL]
  {10.3847/2041-8213/aa9d81}, 851, L31

\bibitem[\protect\citeauthoryear{Laughlin \& Batygin}{Laughlin \&
  Batygin}{2017}]{Laughlin+2017}
Laughlin G.,  Batygin K.,  2017, \mn@doi [RNAAS] {10.3847/2515-5172/aaa02b}, 1,
  43

\bibitem[\protect\citeauthoryear{Levine, Taylor, Seligman, Hoover, Jedicke,
  Bergner  \& Laughlin}{Levine et~al.}{2023}]{Levine+2023}
Levine W.~G.,  Taylor A.~G.,  Seligman D.~Z.,  Hoover D.~J.,  Jedicke R.,
  Bergner J.~B.,   Laughlin G.~P.,  2023, \mn@doi [PSJ] {10.3847/PSJ/acdf58},
  4, 124

\bibitem[\protect\citeauthoryear{Mashchenko}{Mashchenko}{2019}]{Mashchenko+2019}
Mashchenko S.,  2019, \mn@doi [MNRAS] {10.1093/mnras/stz2380}, 489, 3003

\bibitem[\protect\citeauthoryear{McNeill, Trilling  \& Mommert}{McNeill
  et~al.}{2018}]{McNeill+2018}
McNeill A.,  Trilling D.~E.,   Mommert M.,  2018, \mn@doi [ApJL]
  {10.3847/2041-8213/aab9ab}, 857, L1

\bibitem[\protect\citeauthoryear{Meech et~al.,}{Meech
  et~al.}{2017}]{Meech+2017}
Meech K.~J.,  et~al., 2017, \mn@doi [Nature]
  {https://doi.org/10.1038/nature25020}, 552, 378

\bibitem[\protect\citeauthoryear{Micheli et~al.,}{Micheli
  et~al.}{2018}]{Micheli+2018}
Micheli M.,  et~al., 2018, \mn@doi [Nature]
  {https://doi.org/10.1038/s41586-018-0254-4}, 559, 223

\bibitem[\protect\citeauthoryear{Mulders, Pascucci  \& Apai}{Mulders
  et~al.}{2015}]{Mulders+2015}
Mulders G.~D.,  Pascucci I.,   Apai D.,  2015, \mn@doi [ApJ]
  {10.1088/0004-637X/814/2/130}, 814, 130

\bibitem[\protect\citeauthoryear{Petrovich}{Petrovich}{2015}]{Petrovich+2015}
Petrovich C.,  2015, \mn@doi [ApJ] {10.1088/0004-637X/808/2/120}, 808, 120

\bibitem[\protect\citeauthoryear{Pfalzner, Vargas, Bhandare  \& Veras}{Pfalzner
  et~al.}{2021}]{Pfalzner+2021}
Pfalzner S.,  Vargas L. L.~A.,  Bhandare A.,   Veras D.,  2021, \mn@doi [A\&A]
  {https://doi.org/10.1051/0004-6361/202140587}, 651, A38

\bibitem[\protect\citeauthoryear{Quirrenbach et~al.,}{Quirrenbach
  et~al.}{2022}]{Quirrenbach+2022}
Quirrenbach A.,  et~al., 2022, \mn@doi [A\&A]
  {https://doi.org/10.1051/0004-6361/202142915}, 663, A48

\bibitem[\protect\citeauthoryear{Raymond, O’Brien, Morbidelli  \&
  Kaib}{Raymond et~al.}{2009}]{Raymond+2009}
Raymond S.~N.,  O’Brien D.~P.,  Morbidelli A.,   Kaib N.~A.,  2009, \mn@doi
  [Icarus] {10.1016/j.icarus.2009.05.016}, 203, 644

\bibitem[\protect\citeauthoryear{Raymond et~al.,}{Raymond
  et~al.}{2012}]{Raymond+2012}
Raymond S.~N.,  et~al., 2012, \mn@doi [A\&A]
  {https://doi.org/10.1051/0004-6361/201117049}, 541, A11

\bibitem[\protect\citeauthoryear{Raymond, Armitage  \& Veras}{Raymond
  et~al.}{2018}]{Raymond+2018}
Raymond S.~N.,  Armitage P.~J.,   Veras D.,  2018, \mn@doi [ApJL]
  {10.3847/2041-8213/aab4f6}, 856, L7

\bibitem[\protect\citeauthoryear{Raymond, Kaib, Armitage  \& Fortney}{Raymond
  et~al.}{2020}]{Raymond+2020}
Raymond S.~N.,  Kaib N.~A.,  Armitage P.~J.,   Fortney J.~J.,  2020, \mn@doi
  [ApJL] {10.3847/2041-8213/abc55f}, 904, L4

\bibitem[\protect\citeauthoryear{Rein \& Liu}{Rein \& Liu}{2012}]{Rein+2012}
Rein H.,  Liu S.~F.,  2012, \mn@doi [A\&A] {10.1051/0004-6361/201118085}, 537,
  A128

\bibitem[\protect\citeauthoryear{Richardson, Bottke  \& Love}{Richardson
  et~al.}{1998}]{Richardson+1998}
Richardson D.~C.,  Bottke W.~F.,   Love S.~G.,  1998, \mn@doi [Icarus]
  {https://doi.org/10.1006/icar.1998.5954}, 134, 47

\bibitem[\protect\citeauthoryear{Rodet \& Lai}{Rodet \& Lai}{2023}]{Rodet+2023}
Rodet L.,  Lai D.,  2023, \mn@doi [MNRAS] {10.1093/mnras/stad3905}, 527, 11664

\bibitem[\protect\citeauthoryear{Safronov}{Safronov}{1972}]{Safronov+1972}
Safronov V.~S.,  1972, Evolution of the protoplanetary cloud and formation of
  the earth and the planets.
\url
  {https://ia600208.us.archive.org/21/items/nasa_techdoc_19720019068/19720019068.pdf}

\bibitem[\protect\citeauthoryear{Sridhar \& Tremaine}{Sridhar \&
  Tremaine}{1992}]{Sridhar+1992}
Sridhar S.,  Tremaine S.,  1992, \mn@doi [Icarus]
  {https://doi.org/10.1016/0019-1035(92)90193-B}, 95, 86

\bibitem[\protect\citeauthoryear{}{Tea}{2019}]{Team+2019}
 `Oumuamua ISSI Team 2019, \mn@doi [Nat Astron]
  {https://doi.org/10.1038/s41550-019-0816-x}, 7, 594

\bibitem[\protect\citeauthoryear{Udry \& Santos}{Udry \&
  Santos}{2007}]{Udry+2007}
Udry S.,  Santos N.~C.,  2007, \mn@doi [ARA\&A]
  {10.1146/annurev.astro.45.051806.110529}, 45, 397

\bibitem[\protect\citeauthoryear{Walsh}{Walsh}{2018}]{Walsh+2018}
Walsh K.~J.,  2018, \mn@doi [ARA\&A]
  {https://doi.org/10.1146/annurev-astro-081817-052013}, 56, 593

\bibitem[\protect\citeauthoryear{Warner, b  \& Pravec}{Warner
  et~al.}{2009}]{Warner+2009}
Warner B.~D.,  b A. W.~H.,   Pravec P.,  2009, \mn@doi [Icarus]
  {10.1016/j.icarus.2009.02.003}, 202, 134

\bibitem[\protect\citeauthoryear{Weaver et~al.,}{Weaver
  et~al.}{1995}]{Weaver+1995}
Weaver H.~A.,  et~al., 1995, \mn@doi [Science] {10.1126/science.7871424}, 267,
  1282

\bibitem[\protect\citeauthoryear{Winn \& Fabrycky}{Winn \&
  Fabrycky}{2015}]{Winn+2015}
Winn J.~N.,  Fabrycky D.~C.,  2015, \mn@doi [ARA\&A]
  {https://doi.org/10.1146/annurev-astro-082214-122246}, 53, 409

\bibitem[\protect\citeauthoryear{Zhang \& Lin}{Zhang \& Lin}{2020}]{Zhang+2020}
Zhang Y.,  Lin D. N.~C.,  2020, \mn@doi [Nat Astron]
  {https://doi.org/10.1038/s41550-020-1065-8}, 4, 852

\bibitem[\protect\citeauthoryear{Zhou \& Lin}{Zhou \& Lin}{2007}]{Zhou+2007}
Zhou J.~L.,  Lin D. N.~C.,  2007, \mn@doi [ApJ] {10.1086/520043}, 666, 447

\bibitem[\protect\citeauthoryear{Zhou, Aarseth, Lin  \& Nagasawa}{Zhou
  et~al.}{2005}]{Zhou+2005}
Zhou J.~L.,  Aarseth S.~J.,  Lin D. N.~C.,   Nagasawa M.,  2005, \mn@doi [ApJ]
  {10.1086/497094}, 631, L85

\bibitem[\protect\citeauthoryear{Zhu \& Dong}{Zhu \& Dong}{2021}]{Zhu+2021}
Zhu W.,  Dong S.,  2021, \mn@doi [ARA\&A]
  {https://doi.org/10.1146/annurev-astro-112420-020055}, 59, 291

\bibitem[\protect\citeauthoryear{Zuluaga, Sánchez-Hernández, Sucerquia  \&
  Ferrín}{Zuluaga et~al.}{2018}]{Zuluaga+2018}
Zuluaga J.~I.,  Sánchez-Hernández O.,  Sucerquia M.,   Ferrín I.,  2018,
  \mn@doi [AJ] {https://doi.org/10.3847/1538-3881/aabd7c}, 155, 236

\bibitem[\protect\citeauthoryear{Zwart, Torres, Pelupessy, Bédorf  \&
  Cai}{Zwart et~al.}{2018}]{Zwart+2018}
Zwart P.~S.,  Torres S.,  Pelupessy I.,  Bédorf J.,   Cai M.~X.,  2018,
  \mn@doi [MNRASL] {https://doi.org/10.1093/mnrasl/sly088}, 479, L17

\bibitem[\protect\citeauthoryear{\v{D}urech et~al.,}{\v{D}urech
  et~al.}{2012}]{Durech+2012}
\v{D}urech J.,  et~al., 2012, \mn@doi [A\&A] {10.1051/0004-6361/201219396},
  547, A10

\makeatother
\end{thebibliography}




\appendix

\section{The Detailed Results of Each Simulated System}

\begin{table*}
  \begin{center}
    \begin{tabular}{c|c|c|c|c|c|c|c|c|c|c} 
    \hline
    System & $a_\text{p}/\text{au}$ & $e_\text{p}$ & $\Theta$ & $P(\text{ej})$ & $P(\text{tidal}|\text{ej})$ & $P(\text{co-p})$ & $P(\text{co-s})$ & $P(\text{su})$\\
    \hline
    Low mass star-Neptune & $0.026$ & $0.2$ &  $0.0278$  &$1.78\%$& $0.00\%$& $67.36\%$ & $0.54\%$& $30.31\%$\\
    Sun-Jupiter & $5.2$ & $0.048$ &  $11.1$  & $40.69\%$& $0.00\%$ & $11.99\%$ & $0.10\%$& $47.23\%$\\
    Low mass star-Jupiter & $0.026$ & $0.2$ & $0.106$  &$15.53\%$& $0.83\%\pm0.10\%$ & $67.85\%$ & $3.52\%$& $13.09\%$\\
    Low mass star-Jupiter & $0.052$ & $0.2$ & $0.212$ &$20.45\%$& $0.89\%\pm0.09\%$ & $62.84\%$ & $3.54\%$& $13.17\%$\\
    Low mass star-Jupiter & $0.104$ & $0.2$ & $0.423$ &$28.41\%$& $0.72\%\pm0.07\%$& $54.11\%$ & $4.27\%$& $13.21\%$\\
    Low mass star-Jupiter & $0.26$ & $0.2$ & $1.06$ &$42.13\%$& $0.72\%\pm0.06\%$&$39.21\%$ & $5.49\%$& $13.18\%$\\
    Low mass star-Jupiter & $0.52$ & $0.2$ & $2.12$ &$52.48\%$& $0.62\%\pm0.05\%$& $28.50\%$ & $5.86\%$& $13.16\%$\\
    Low mass star-Jupiter & $1.04$ & $0.2$ &  $4.23$ &$61.52\%$& $0.54\%\pm0.04\%$& $18.87\%$ & $6.35\%$& $13.26\%$\\
    Low mass star-Jupiter & $2.6$ & $0.2$ & $10.6$ &$70.65\%$& $0.55\%\pm0.04\%$& $9.68\%$ & $6.37\%$& $13.30\%$\\
    Low mass star-Jupiter & $5.2$ & $0.2$ &  $21.1$  &$75.09\%$& $0.61\%\pm0.04\%$& $5.68\%$ & $5.98\%$& $13.25\%$\\
    Low mass star-Jupiter & $10.4$ & $0.2$ &  $42.3$ &$78.21\%$& $0.72\%\pm0.04\%$& $3.22\%$ & $5.29\%$& $13.29\%$\\
    Low mass star-Jupiter & $26$ & $0.2$ & $106$ &$80.59\%$& $0.66\%\pm0.04\%$& $1.34\%$ & $4.78\%$& $13.29\%$\\
    \hline
    \end{tabular}
  \end{center}
  \caption{The settings and results (the probability of a planetesimal in the chaotic zone evolved to be ejected $P(\text{ej})$, and the probability of an ejected planetesimal undergoing tidal fragmentation $P(\text{tidal}|\text{ej})$, the probability to collide with the planet $P(\text{co-p})$, with the star $P(\text{co-s})$ and survive $P(\text{su})$) of each system.}
  \label{table1}
\end{table*}

\begin{table*}
  \begin{center}
    \begin{tabular}{c|c|c|c|c|c|c|c|c|c|c|c|c} 
    \hline
    $M_\text{p}$ & $a_\text{in}/\text{au}$ & $a_\text{out}/\text{au}$ & $P(\text{ej})$ & $P(\text{tidal}|\text{ej})$& $P(\text{co-p})$ & $P(\text{co-s})$ & $P(\text{su})$ \\
    \hline
    $M_\text{J}$ & $5.30$ & $19.45$ &$77.70\%$& $0.73\%\pm0.04\%$& $3.04\%$ & $5.32\%$& $13.94\%$\\
    $2M_\text{J}$ & $4.89$ & $21.01$ &$77.79\%$& $0.51\%\pm0.04\%$& $2.86\%$ & $4.20\%$& $15.15\%$\\
    $4M_\text{J}$ & $4.47$ & $23.09$ &$79.08\%$& $0.25\%\pm0.03\%$& $2.51\%$ & $4.10\%$& $14.31\%$\\
    \hline
    \end{tabular}
  \end{center}
  \caption{Same as Table \ref{table1}, with different planetary masses and boundaries [$a_\text{in}, a_\text{out}$]. Planetary semi-major axes and eccentricities are taken as $a_\text{p}=2a_\text{J}=10.4\,\text{au}$, $e_\text{p}=0.2$.}
  \label{table2}
\end{table*}

\begin{table*}
  \begin{center}
\begin{tabular}{c|c|c|c|c|c|c|c|c|c|c} 
    \hline
    $a_\text{p}/\text{au}$ & $d_{\text{str-s}}/\text{au}$ & $d_{\text{str-p}}/\text{au}$ & $P(\text{tidal-s}|\text{ej})$ & $P(\text{tidal-p}|\text{ej})$ & $P(\text{tidal-s\&p}|\text{ej})$ \\
    \hline
    $0.026$ & $0.162$ & $0.0201$ & $0.83\%\pm0.10\%$ & $3.43\%\pm0.20\%$ & $0.16\%\pm0.04\%$ \\
    $0.052$ & $0.0810$ & $0.0100$ & $0.89\%\pm0.09\%$ & $4.66\%\pm0.20\%$ & $0.22\%\pm0.04\%$ \\
    $0.104$ & $0.0405$ & $0.00502$ & $0.72\%\pm0.07\%$ & $5.14\%\pm0.18\%$ & $0.10\%\pm0.03\%$ \\
    $0.26$ & $0.0162$ & $0.00201$ & $0.72\%\pm0.06\%$ & $3.84\%\pm0.13\%$ & $0.08\%\pm0.02\%$ \\
    $0.52$ & $0.00810$ & $1.00\times10^{-3}$ & $0.62\%\pm0.05\%$ & $2.74\%\pm0.10\%$ & $0.02\%\pm0.01\%$ \\
    $1.04$ & $0.00405$ & $5.02\times10^{-4}$ & $0.54\%\pm0.04\%$ & $1.98\%\pm0.08\%$ & $0.00\%$ \\
    $2.6$ & $0.00162$ & $2.01\times10^{-4}$ & $0.55\%\pm0.04\%$ & $1.24\%\pm0.06\%$ & $0.00\%$ \\
    $5.2$ & $8.10\times10^{-4}$ & $1.00\times10^{-4}$ & $0.61\%\pm0.04\%$ & $0.64\%\pm0.04\%$ & $0.00\%$ \\
    $10.4$ & $4.05\times10^{-4}$ & $5.02\times10^{-5}$ & $0.72\%\pm0.04\%$ & $0.39\%\pm0.04\%$ & $0.00\%$ \\
    $26$ & $1.62\times10^{-4}$ & $2.01\times10^{-5}$ & $0.66\%\pm0.04\%$ & $0.16\%\pm0.04\%$ & $0.00\%$ \\
    \hline
\end{tabular}
\end{center}
\caption{Under the settings of $\phi=35^{\circ}$, $M_{\text{s}}=0.5M_\odot$, $R_{\text{s}}=0.5R_\odot$, $M_{\text{p}}=M_\text{J}$, $R_{\text{p}}=R_\text{J}$, $e_{\text{p}}=0.2$, $d_{\text{str-s}}=4.21\times10^{-3}\,\text{au}=6.30\times10^{8}\,\text{m}$, $d_{\text{str-p}}=5.20\times10^{-4}\,\text{au}=7.78\times10^{7}\,\text{m}$, the probability of an ejected planetesimal tidally fragmented by the star $P(\text{tidal-s}|\text{ej})$, by the planet $P(\text{tidal-p}|\text{ej})$, and both $P(\text{tidal-s\&p}|\text{ej})$ of each system with different planetary semi-major axes.}
\label{table3}
\end{table*}

\begin{table*}
  \begin{center}
\begin{tabular}{c|c|c|c|c|c|c|c|c|c|c} 
    \hline
    $a_\text{p}/\text{au}$ & $d_{\text{str-s}}/\text{au}$ & $d_{\text{str-p}}/\text{au}$ & $P(\text{tidal-s}|\text{ej})$ & $P(\text{tidal-p}|\text{ej})$ & $P(\text{tidal-s\&p}|\text{ej})$ \\
    \hline
    $0.026$ & $0.171$ & $0.0212$ & $0.91\%\pm0.10\%$ & $4.75\%\pm0.23\%$ & $0.23\%\pm0.05\%$ \\
    $0.052$ & $0.0857$ & $0.0106$ & $0.96\%\pm0.09\%$ & $7.14\%\pm0.25\%$ & $0.32\%\pm0.05\%$ \\
    $0.104$ & $0.0428$ & $0.00531$ & $0.81\%\pm0.07\%$ & $7.51\%\pm0.21\%$ & $0.17\%\pm0.03\%$ \\
    $0.26$ & $0.0171$ & $0.00212$ & $0.77\%\pm0.06\%$ & $5.88\%\pm0.16\%$ & $0.11\%\pm0.02\%$ \\
    $0.52$ & $0.00857$ & $1.06\times10^{-3}$ & $0.71\%\pm0.05\%$ & $4.28\%\pm0.12\%$ & $0.05\%\pm0.01\%$ \\
    $1.04$ & $0.00428$ & $5.31\times10^{-4}$ & $0.57\%\pm0.04\%$ & $2.99\%\pm0.09\%$ & $0.01\%\pm0.01\%$ \\
    $2.6$ & $0.00171$ & $2.12\times10^{-4}$ & $0.60\%\pm0.04\%$ & $1.88\%\pm0.07\%$ & $0.01\%\pm0.00\%$ \\
    $5.2$ & $8.57\times10^{-4}$ & $1.06\times10^{-4}$ & $0.67\%\pm0.04\%$ & $1.03\%\pm0.05\%$ & $0.01\%\pm0.00\%$ \\
    $10.4$ & $4.28\times10^{-4}$ & $5.31\times10^{-5}$ & $0.78\%\pm0.04\%$ & $0.64\%\pm0.04\%$ & $0.01\%\pm0.00\%$ \\
    $26$ & $1.71\times10^{-4}$ & $2.12\times10^{-5}$ & $0.72\%\pm0.04\%$ & $0.25\%\pm0.02\%$ & $0.00\%$ \\
    \hline
\end{tabular}
  \end{center}
  \caption{Under the settings of $\phi=30^{\circ}$, $M_\text{s}=0.5M_\odot$, $R_\text{s}=0.5R_\odot$, $M_\text{p}=M_\text{J}$, $R_\text{p}=R_\text{J}$, $e_{\text{p}}=0.2$, $d_{\text{str-s}}=4.46\times10^{-3}\,\text{au}=6.67\times10^{8}\,\text{m}$, $d_{\text{str-p}}=5.51\times10^{-4}\,\text{au}=8.25\times10^{7}\,\text{m}$, the probability of an ejected planetesimal tidally fragmented by the star $P(\text{tidal-s}|\text{ej})$, by the planet $P(\text{tidal-p}|\text{ej})$, and both $P(\text{tidal-s\&p}|\text{ej})$ of each system with different planetary semi-major axes.}
  \label{table4}
\end{table*}

\begin{table*}
  \begin{center}
\begin{tabular}{c|c|c|c|c|c|c|c|c|c|c} 
    \hline
    $a_\text{p}/\text{au}$ & $d_{\text{str-s}}/\text{au}$ & $d_{\text{str-p}}/\text{au}$ & $P(\text{tidal-s}|\text{ej})$ & $P(\text{tidal-p}|\text{ej})$ & $P(\text{tidal-s\&p}|\text{ej})$ \\
    \hline
    $0.026$ & $0.183$ & $0.0227$ & $1.11\%\pm0.11\%$ & $6.36\%\pm0.27\%$ & $0.42\%\pm0.07\%$ \\
    $0.052$ & $0.0915$ & $0.0114$ & $1.15\%\pm0.10\%$ & $10.21\%\pm0.29\%$ & $0.56\%\pm0.07\%$ \\
    $0.104$ & $0.0458$ & $0.00568$ & $0.93\%\pm0.08\%$ & $10.36\%\pm0.25\%$ & $0.28\%\pm0.04\%$ \\
    $0.26$ & $0.0183$ & $0.00227$ & $0.80\%\pm0.06\%$ & $8.30\%\pm0.18\%$ & $0.17\%\pm0.03\%$ \\
    $0.52$ & $0.00915$ & $1.14\times10^{-3}$ & $0.77\%\pm0.05\%$ & $6.13\%\pm0.14\%$ & $0.08\%\pm0.02\%$ \\
    $1.04$ & $0.00458$ & $5.68\times10^{-4}$ & $0.61\%\pm0.04\%$ & $4.32\%\pm0.11\%$ & $0.02\%\pm0.01\%$ \\
    $2.6$ & $0.00183$ & $2.27\times10^{-4}$ & $0.65\%\pm0.04\%$ & $2.71\%\pm0.08\%$ & $0.01\%\pm0.01\%$ \\
    $5.2$ & $9.15\times10^{-4}$ & $1.14\times10^{-4}$ & $0.73\%\pm0.04\%$ & $1.56\%\pm0.06\%$ & $0.01\%\pm0.01\%$ \\
    $10.4$ & $4.58\times10^{-4}$ & $5.68\times10^{-5}$ & $0.85\%\pm0.04\%$ & $0.92\%\pm0.05\%$ & $0.01\%\pm0.01\%$ \\
    $26$ & $1.83\times10^{-4}$ & $2.27\times10^{-5}$ & $0.82\%\pm0.04\%$ & $0.36\%\pm0.03\%$ & $0.00\%$ \\
    \hline
\end{tabular}
\end{center}
\caption{Under the settings of $\phi=25^{\circ}$, $M_\text{s}=0.5M_\odot$, $R_\text{s}=0.5R_\odot$, $M_\text{p}=M_\text{J}$, $R_\text{p}=R_\text{J}$, $e_{\text{p}}=0.2$, $d_{\text{str-s}}=4.76\times10^{-3}\,\text{au}=7.12\times10^{8}\,\text{m}$, $d_{\text{str-p}}=5.93\times10^{-4}\,\text{au}=8.87\times10^{7}\,\text{m}$, the probability of an ejected planetesimal tidally fragmented by the star $P(\text{tidal-s}|\text{ej})$, by the planet $P(\text{tidal-p}|\text{ej})$, and both $P(\text{tidal-s\&p}|\text{ej})$ of each system with different planetary semi-major axes.}
\label{table5}
\end{table*}


\bsp	
\label{lastpage}
\end{document}